\documentclass{epl}

\title{Event-based simulation of single-photon beam splitters \\
and Mach-Zehnder interferometers%
\footnote{
Email: $^1$deraedt@phys.rug.nl; $^2$deraedt@cs.rug.nl; $^3$kristel@phys.rug.nl\\
Homepage: //www.compphys.org}
}
\author{H. De Raedt\inst{1}, K. De Raedt\inst{2} \and K. Michielsen\inst{3}}
\institute{
  \inst{1}
Department of Applied Physics,
Materials Science Centre, University of Groningen, Nijenborgh 4,
NL-9747 AG Groningen, The Netherlands \\
  \inst{2}
Department of Computer Science,
University of Groningen, Blauwborgje 3,
NL-9747 AC Groningen, The Netherlands
}
\begin{document}
\maketitle

\begin{abstract}
We demonstrate that networks of locally connected
processing units with a primitive learning capability
exhibit behavior that is usually only attributed to quantum systems.
We describe networks that simulate
single-photon beam-splitter and Mach-Zehnder interferometer experiments
on a causal, event-by-event basis and demonstrate that
the simulation results are in excellent agreement with quantum theory.
\end{abstract}

\def\ORDER#1{\hbox{${\cal O}(#1)$}}
\def\BRA#1{\langle #1 \vert}
\def\KET#1{\vert #1 \rangle}
\def\EXPECT#1{\langle #1 \rangle}
\def\BRACKET#1#2{\langle #1 \vert #2 \rangle}
\def\hbar{{\mathchar'26\mskip-9muh}}
\def\mod{{\mathop{\hbox{mod}}}}
\def\CNOT{{\mathop{\hbox{CNOT}}}}
\def\Tr{{\mathop{\hbox{Tr}}}}
\def\bPsi{{\mathbf{\Psi}}}
\def\bPhi{{\mathbf{\Phi}}}
\def\bzero{{\mathbf{0}}}
\def\Eq#1{(\ref{#1})}

\def\DLM{SLM}
\def\DLMS{SLMs}

\section{Introduction}\label{intro}

The objective of the research reported in this paper is to
demonstrate that locally-connected networks of processing units
with a primitive learning capability are sufficient
to simulate, event-by-event, the single-photon beam splitter and Mach-Zehnder interferometer
experiments of Grangier et al.~\cite{GRAN86}.
This is one of the basic experiments in quantum physics~\cite{GRAN86,TONO98}
that has not been simulated in the event-by-event manner in which the experimental observations
are actually recorded~\cite{PerfectExperiments}.
Although quantum theory gives us a recipe to compute the frequency for
observing different types of events,
it does not describe individual events~\cite{BALL03,HOME97,TONO98}.
Reconciling the mathematical formalism (that does not describe single events) with
the experimental fact that each observation yields a definite
outcome is often referred to as the quantum measurement paradox.
This is a central, fundamental problem in the foundation
of quantum theory~\cite{FEYN65,HOME97,PENR90}.

From a computational viewpoint, quantum theory provides us
with a set of rules (algorithms) to compute
probability distributions~\cite{HOME97,KAMP88,QuantumTheory}.
Therefore, we may wonder what kind of algorithm(s) we need to perform an event-based
simulation of the experiments mentioned above without using wave functions.
Evidently, this formulation rules out any method based on the solution of the
(time-dependent) Schr{\"o}dinger equation.
We have to step outside the framework that quantum theory provides.

We formulate the simulation of quantum processes in terms of events, messages,
and units that process these events and messages.
In terms of the experiments of Grangier et al.~\cite{GRAN86},
the photon carries the message (a phase),
an event is the arrival of a photon at one of the input ports
of a beam splitter, and the beam splitters are the processing units.
The essential feature of a processing unit is its ability
to learn from the events it processes.
We use a standard linear adaptive filter~\cite{HAYK86} for this purpose.
The processing unit sends a message (carried by a photon)
through an output port that is selected randomly
according to a distribution that is determined
by the current state of the processing unit.
A processing unit that operates according to this principle
will be referred to as a stochastic learning machine (SLM).
Here the term {\sl stochastic} does not refer to the learning process but to
the method that is used to select the output channel that will carry the outgoing message.
The learning process itself is deterministic.
Elsewhere we describe processing units that are fully deterministic
and are not based on linear adaptive filters, but
can be used for exactly the same purpose and also
for simulating many-body quantum phenomena~\cite{KOEN04,HANS05,KRIS04,MZIdemo}.

By connecting an output channel to the input channel of another \DLM\ we can build
networks of \DLMS.
As the input of a network receives an event,
the corresponding message is routed through the network while it is being processed.
At any given time during the processing,
there is only one input-output connection in the network that is actually carrying a message.
The \DLMS\ process the messages in a sequential manner and communicate with each other
by message passing.
There is no other form of communication between different \DLMS.
Although networks of \DLMS\ can be viewed as networks that are capable of unsupervised learning,
they have little in common with neural networks~\cite{HAYK99}.

\section{Beam splitter}\label{BS}\label{ILLUa}\label{QI}\label{ILLU}\label{HYP}

\begin{figure}
\begin{center}
\includegraphics[width=12cm]{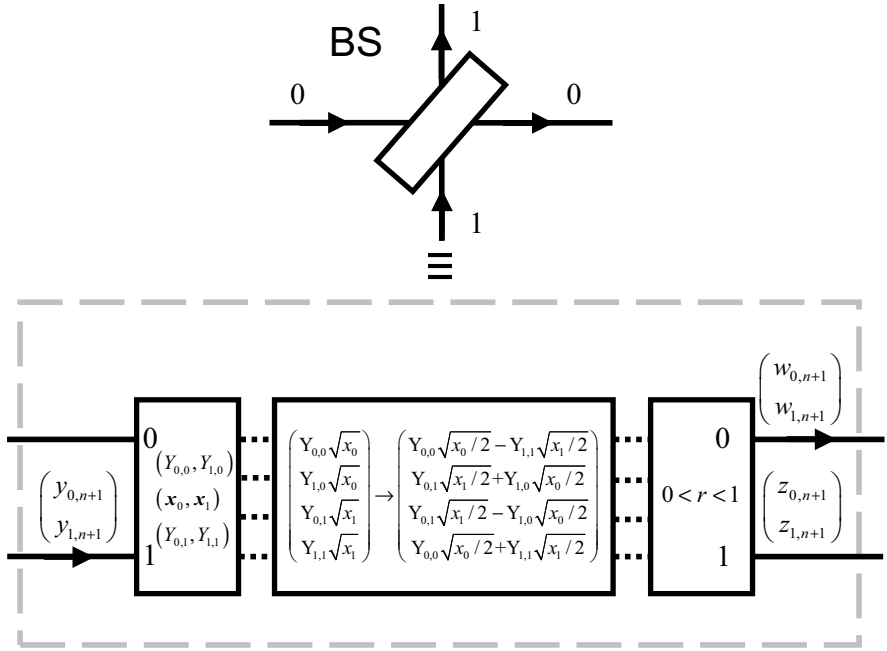}
\caption{
Diagram of a \DLM\ that performs an event-by-event
simulation of a single-photon beam splitter (BS).
The solid lines represent the input and output channels of the BS.
Dashed lines indicate the flow of data within the BS.
}
\label{figbs}
\end{center}
\end{figure}

According to quantum theory~\cite{QuantumTheory}, the amplitudes ($b_0,b_1)$
of the photons in the output modes 0 and 1 of a
beam splitter (see Fig.~\ref{figbs}) are given by
\begin{eqnarray}
\left(
\begin{array}{c}
b_0\\
b_1
\end{array}
\right)
=
\frac{1}{\sqrt{2}}
\left(
\begin{array}{c}
a_0+ia_1\\
a_1+ia_0
\end{array}
\right)
=
\frac{1}{\sqrt{2}}
\left(
\begin{array}{cc}
1&i\\
i&1
\end{array}
\right)
\left(
\begin{array}{c}
a_0\\
a_1
\end{array}
\right),
\label{BS3}
\end{eqnarray}
where the presence of photons in the input modes 0 or 1 is represented
by the amplitudes ($a_0,a_1)$~\cite{BAYM74,GRAN86,RARI97,NIEL00}.
Here we construct a \DLM\ that acts as a
beam splitter, not by calculating the amplitudes according to Eq.~\Eq{BS3} but
by processing individual events (photons).
First we describe the various components of the machine.
Then we demonstrate that it acts as a beam splitter.

A schematic diagram of the \DLM\ is shown in Fig.~\ref{figbs}.
We label events by a subscript $n\ge0$.
At the $(n+1)$-th event, the \DLM\ receives a message on either input channel
0 or 1, never on both channels simultaneously.
If the event occurs on channel 0 (1) we call it an event of type 0 (1).
Every message is a two-dimensional unit vector ${\bf y}_{n+1}=(y_{0,n+1},y_{1,n+1})$.
This message represents the time-of-flight, that is the length of the optical
path of the photon when it travels from source to beam splitter, from beam splitter
to beam splitter and the like.
In quantum theory, this information is encoded in
the phases of the complex numbers $a_0$  and $a_1$.

The first stage (see Fig.~\ref{figbs}) of the \DLM\ (called front-end)
stores the message ${\bf y}_{n+1}$ in its internal register ${\bf Y}_k=(Y_{0,k},Y_{1,k})$
where $k=0$ (1) if the event occurred on channel 0 (1).
The front-end also has an internal two-dimensional vector
${\bf x}=(x_{0},x_{1})$
with the additional constraints that $x_{i}\ge0$ for $i=0,1$
and that $x_{0} +x_{1}=1$.
Here and in Fig.~\ref{figbs}, we have omitted the event-label $n$ from
the internal register ${\bf Y}_k$ and internal vector ${\bf x}$ to simplify
the notation.

After receiving the $(n+1)$-th event on input channel $k=0,1$
the internal vector is updated according to the rule
\begin{eqnarray}
x_{i,n+1}&=&\alpha x_{i,n} + (1-\alpha)\delta_{i,k},
\label{HYP1}
\end{eqnarray}
where $0<\alpha<1$ is a parameter that controls the learning process
and is discussed later.
By construction $x_{i,n+1}\ge0$ for $i=0,1$ and $x_{0,n+1} +x_{1,n+1}=1$.
Hence, the update rule Eq.~\Eq{HYP1} preserves the constraints
on the internal vector.
These constraints are necessary if we want to interpret
the $x_{k,n}$ as (an estimate of) the probability
for the occurrence of an event of type $k$.

\setlength{\unitlength}{1cm}
\begin{figure}[t]
\begin{center}
\begin{picture}(12,5)
\put(-1,0){\includegraphics[width=7cm]{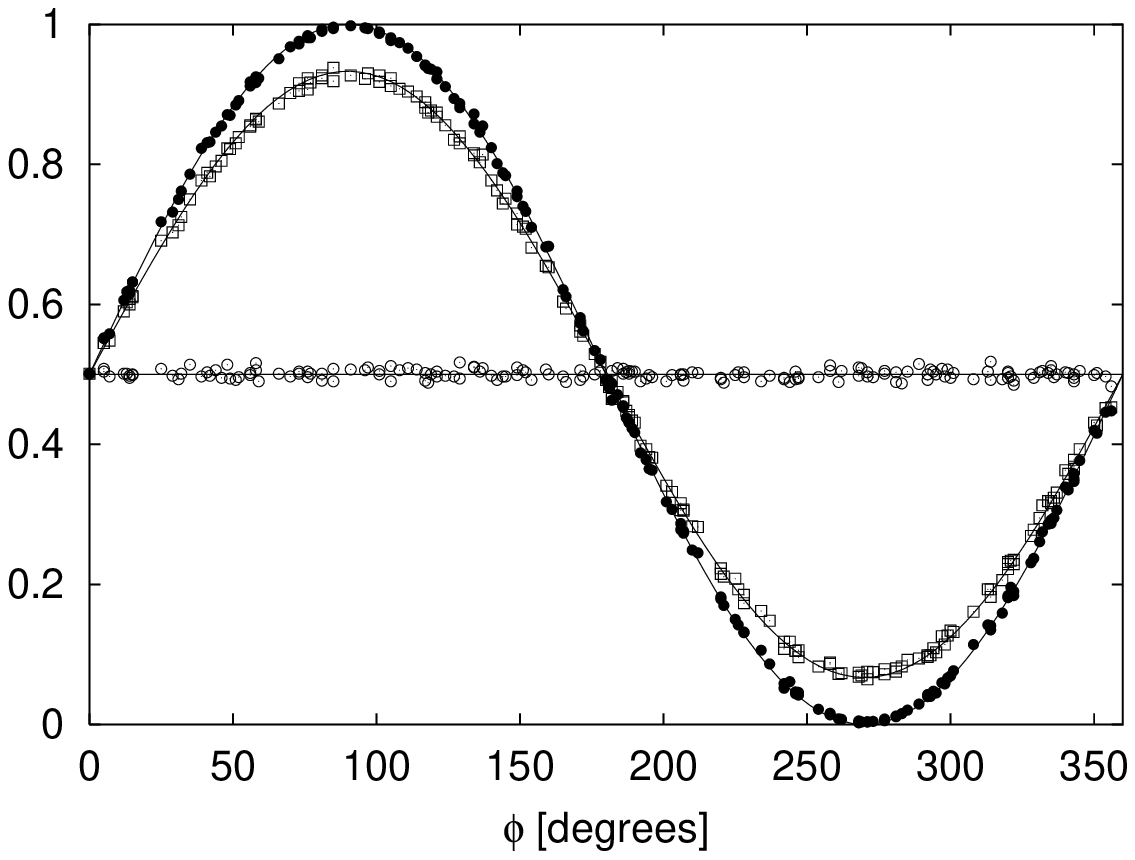}}
\put(6,0){\includegraphics[width=7cm]{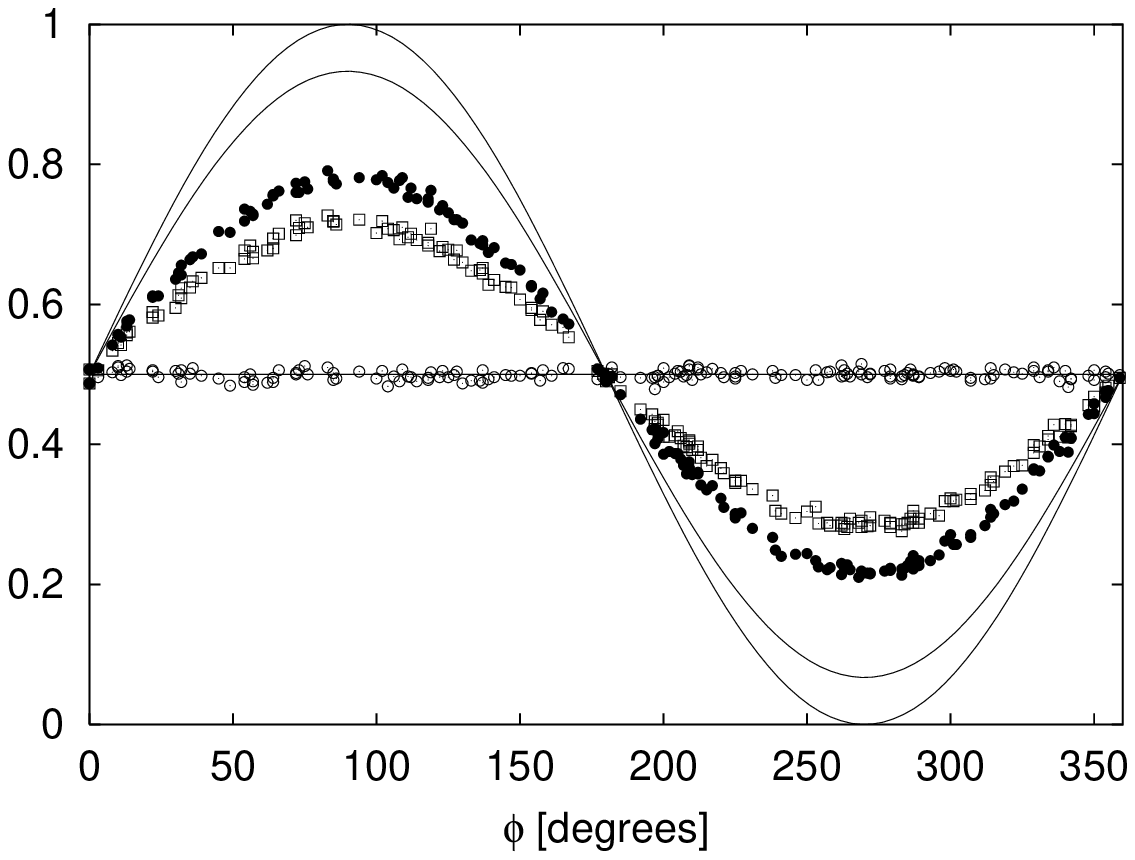}}
\end{picture}
\caption{
Simulation results for the beam splitter shown in Fig.~\ref{figbs}.
Input channel 0 receives $(\cos\psi_0,\sin\psi_0)$
with probability $p_0$.
Input channel 1 receives $(\cos\psi_1,\sin\psi_1)$ with probability
$p_1=1-p_0$.
Each data point represents 10000 events ($N_0+N_1=10000$).
After each set of 10000 events, two uniform
random numbers in the range $[0,360]$ are used to
choose the angles $\psi_0$ and $\psi_1$.
Markers give the simulation results for the
normalized intensity $N_0/(N_0+N_1)$ in output channel 0 as a function of
$\phi=\psi_0-\psi_1$.
Open circles: $p_0=1$;
Bullets: $p_0=0.5$;
Open squares: $p_0=0.25$.
Lines represent the results of quantum theory
(see Eq.~(\ref{BS4})).
Left: $\alpha=0.98$; Right: $\alpha=0.25$.
}
\label{one-bs}
\end{center}
\end{figure}

The solution of Eq.~\Eq{HYP1} reads
\begin{equation}
{\bf x}_{n}=\alpha^n {\bf x}_0 + (1-\alpha)\sum_{i=0}^{n-1}\alpha^{n-1-i}{\bf v}_{i+1}
,
\label{HYP2}
\end{equation}
where ${\bf x}_n=(x_{0,n},x_{1,n})$,
and ${\bf x}_0$ denotes the initial value of the internal vector.
The input events are represented by the vectors
${\bf v}_{n+1}=(1,0)^T$ or ${\bf v}_{n+1}=(0,1)^T$ if the ${n+1}$-th
event occurred on channel 0 or 1, respectively.
Let $p_0$ ($p_1=1-p_0$) be the probability for generating
statistically independent events of type 0 (1)
and let $\langle{\bf x}\rangle_n=({\bf x}_1+\ldots+{\bf x}_n)/n$
denote the Cesaro mean of the sequence $\{{\bf x}_1,\ldots,{\bf x}_n\}$.
Then, if $0<\alpha<1$, we have
$\lim_{n\rightarrow\infty} \langle{\bf x}\rangle_n=(p_0,1-p_0)^T$.
Therefore, the front-end ``learns'' the probabilities
for events 0 and 1 by processing these events in a sequential manner.

The second stage of the \DLM\ in Fig.~\ref{figbs}
takes as input the values stored in the registers ${\bf Y}_0$, ${\bf Y}_1$, {\bf x}
and transforms this data according to the rule
\begin{eqnarray}
\frac{1}{\sqrt{2}}
\left(
\begin{array}{c}
Y_{0,0}\sqrt{x_0}-Y_{1,1}\sqrt{x_1}\\
Y_{0,1}\sqrt{x_1}+Y_{1,0}\sqrt{x_0}\\
Y_{0,1}\sqrt{x_1}-Y_{1,0}\sqrt{x_0}\\
Y_{0,0}\sqrt{x_0}+Y_{1,1}\sqrt{x_1}
\end{array}
\right)
%\overset{BS}{\longrightarrow}
{\longleftarrow}
\left(
\begin{array}{c}
Y_{0,0}\sqrt{x_0}\\
Y_{1,0}\sqrt{x_0}\\
Y_{0,1}\sqrt{x_1}\\
Y_{1,1}\sqrt{x_1}
\end{array}
\right),
\label{BS1}
\end{eqnarray}
where we have omitted the event label $(n+1)$ to simplify the notation.
Using two complex numbers instead of four real numbers
Eq.~\Eq{BS1} can also be written as
\begin{eqnarray}
\frac{1}{\sqrt{2}}
\left(
\begin{array}{c}
Y_{0,0}\sqrt{x_0}-Y_{1,1}\sqrt{x_1}+i(Y_{0,1}\sqrt{x_1}+Y_{1,0}\sqrt{x_0})\\
Y_{0,1}\sqrt{x_1}-Y_{1,0}\sqrt{x_0}+i(Y_{0,0}\sqrt{x_0}+Y_{1,1}\sqrt{x_1})
\end{array}
\right)
%\overset{BS}{\longrightarrow}
{\longleftarrow}
\left(
\begin{array}{c}
Y_{0,0}\sqrt{x_0}+i Y_{1,0}\sqrt{x_0}\\
Y_{0,1}\sqrt{x_1}+i Y_{1,1}\sqrt{x_1}
\end{array}
\right).
\label{BS2}
\end{eqnarray}
Identifying $a_0$ with $Y_{0,0}\sqrt{x_0}+i Y_{1,0}\sqrt{x_0}$
and $a_1$ with $Y_{0,1}\sqrt{x_1}+i Y_{1,1}\sqrt{x_1}$
it is clear that the transformation Eq.\Eq{BS2} plays the
role of the matrix-vector multiplication
in Eq.\Eq{BS3}.

The third stage (see Fig.~\ref{figbs}) of the \DLM\ (called back-end)
responds to the input event by sending a message
${\bf w}_{n+1}=( Y_{0,0}\sqrt{x_0}-Y_{1,1}\sqrt{x_1},Y_{0,1}\sqrt{x_1}+Y_{1,0}\sqrt{x_0})/\sqrt{2}$
through output channel 0 if
$w_{0,n+1}^2+w_{1,n+1}^2<r$
where $0<r<1$ is a uniform random number.
Otherwise, since ${\bf w}_{n+1}^2+{\bf z}_{n+1}^2=1$, the back-end sends the message
${\bf z}_{n+1}=( Y_{0,1}\sqrt{x_1}-Y_{1,0}\sqrt{x_0},Y_{0,0}\sqrt{x_0}+Y_{1,1}\sqrt{x_1})/\sqrt{2}$
through output channel 1.

In Fig.~\ref{one-bs} we present results of event-based simulations using
the \DLM\ depicted in Fig.~\ref{figbs}.
Before the simulation starts we set ${\bf x}_0=(x_{0,0},x_{1,0})=(r,1-r)$
where $r$ is a uniform random number in the interval $[0,1]$.
We also use uniform random numbers to generate two two-dimensional unit vectors
to initialize the registers ${\bf Y}_0$ and ${\bf Y}_1$.
The simulation results of the left and right panel were obtained for $\alpha=0.98$
and $\alpha=0.25$, respectively.
Input channel 0 receives $(\cos\psi_0,\sin\psi_0)$ with probability $p_0$.
Input channel 1 receives $(\cos\psi_1,\sin\psi_1)$ with probability $p_1=1-p_0$.
In the quantum-theory setting, this corresponds to the input amplitudes
$a_0=\sqrt{p_0}e^{i\psi_0}$ and $a_1=\sqrt{1-p_0}e^{i\psi_1}$.
The data in Fig.~\ref{one-bs} show the normalized intensity $N_0/(N_0+N_1)$, where
$N_0$ ($N_1$) denotes the number of events of type 0 (1), as a function of
$\phi=\psi_0-\psi_1$.
Each data point in Fig.~\ref{one-bs} represents a simulation of 10000 events ($N_0+N_1=10000$).
For each set of 10000 events, two uniform
random numbers in the range $[0,360]$ determine the two angles $\psi_0$ and $\psi_1$.
From Fig.~\ref{one-bs}, it is clear that the
\DLM-based beam splitter reproduces the probability distributions
\begin{eqnarray}
|b_0|^2=\frac{1+2\sqrt{p_0(1-p_0)}\sin(\psi_0-\psi_1)}{2},%\nonumber \\
\quad
|b_1|^2=\frac{1-2\sqrt{p_0(1-p_0)}\sin(\psi_0-\psi_1)}{2},
\label{BS4}
\end{eqnarray}
as obtained from Eq.\Eq{BS3}.

\section{Role of the control parameter $\alpha$}\label{ALPHA}

From Eq.~(\ref{HYP1}), we see that $\alpha$ controls the speed of learning.
Furthermore, it is evident that the difference between
a constant input to a \DLM\ and the learned value of its
internal variable cannot be smaller than $1-\alpha$.
In other words, $\alpha$ also limits the precision with
which the internal variable can represent a sequence
of constant input values.
On the other hand, the number of events $N$ has to balance
the rate at which the \DLM\ can forget a learned input value.
The smaller $1-\alpha$ is, the larger the number of events
has to be for the \DLM\ to adapt to changes in the input data.
Simulations (results not shown) show that the error decreases
as $1/\sqrt{N}$, as expected from elementary statistics.

In general, for a fixed number of events,
decreasing $\alpha$ leads to an increase of systematic errors.
For instance, if $\alpha=1/4$ the maximum normalized intensity
(at $\phi=90$) in output channel 0 is about 0.8,
as shown in Fig.~\ref{one-bs} (right).
This is easy to understand: In our event-by-event approach,
interference is the result of learning by the SLMs.
If $\alpha$ is close to one, the SLM learns slowly but accurately.
If $\alpha$ decreases, the SLM can adapt faster to changes in the
input data but it also forgets faster.
For very small $\alpha$, ${\bf x}_n$ is always close to
$(1,0)^T$ or $(0,1)^T$ and it becomes impossible to mimic quantum interference.
Thus, $\alpha$ determines the visibility~\cite{GRAN86} of the interference effects.
In our simulation approach, interference is the result of learning (which
requires some form of memory).

\section{Mach-Zehnder interferometer}\label{MZI}

\setlength{\unitlength}{1cm}
\begin{figure}[t]
\begin{center}
\includegraphics[width=8cm]{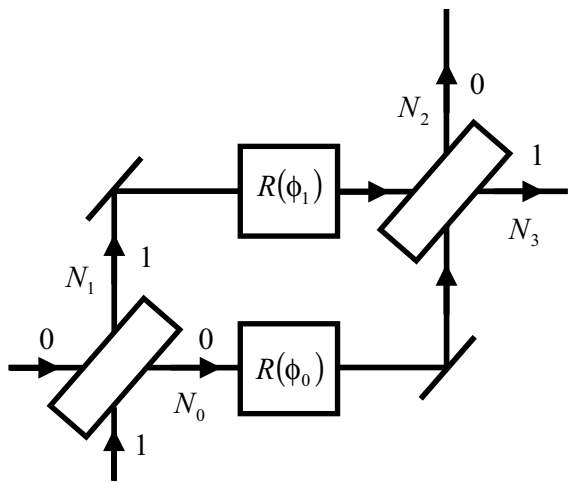}
\caption{
Diagram of a \DLM\ network that simulates a single-photon
Mach-Zehnder interferometer on an event-by-event basis~\cite{MZIdemo}.
The \DLM\ network consists of two BS devices (see Fig.~\ref{figbs})
and two passive devices $R(\phi_0)$ and $R(\phi_1)$
that perform plane rotations by $\phi_0$ and $\phi_1$, respectively.
The number of events $N_i$ in channel $i=0,\ldots,3$
corresponds to the probability for finding a photon
on the corresponding arm of the interferometer.
}
\label{figmz}
\end{center}
\end{figure}

\begin{figure}[t]
\begin{center}
\begin{picture}(12,5)
\put(-1,0){\includegraphics[width=7cm]{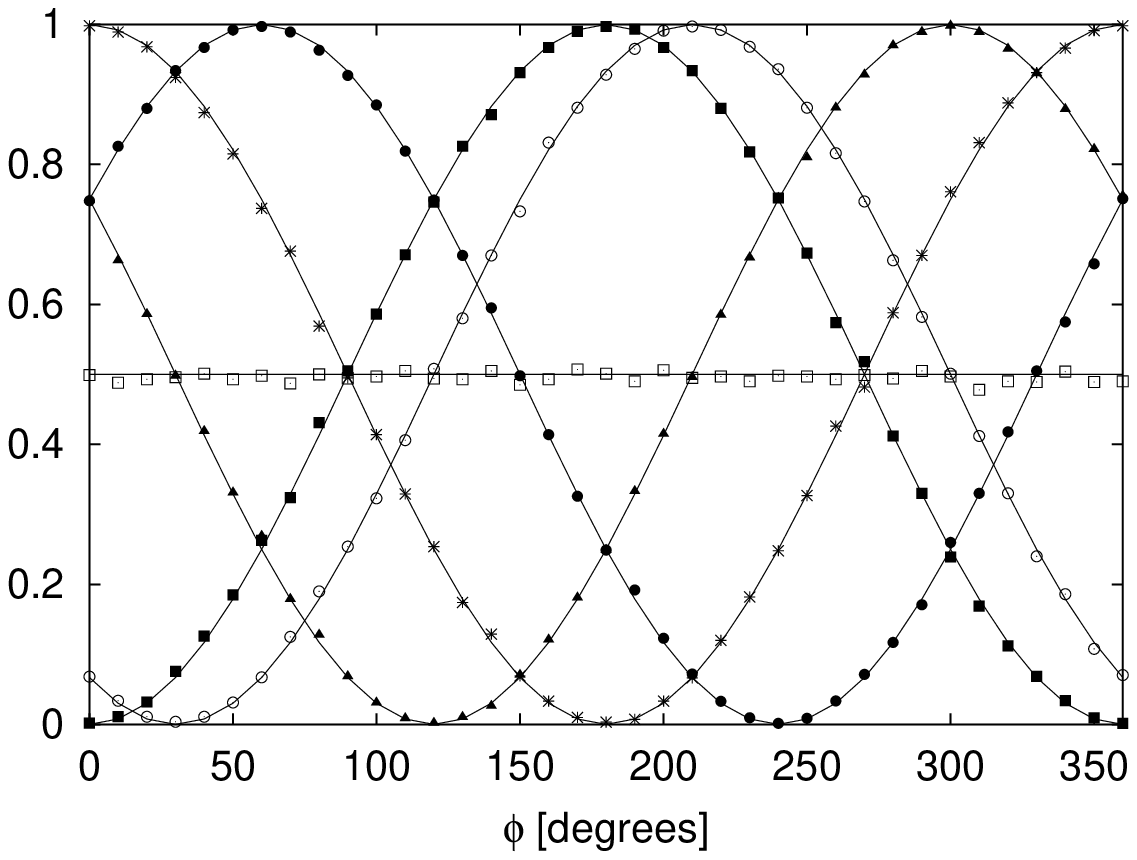}}
\put(6,0){\includegraphics[width=7cm]{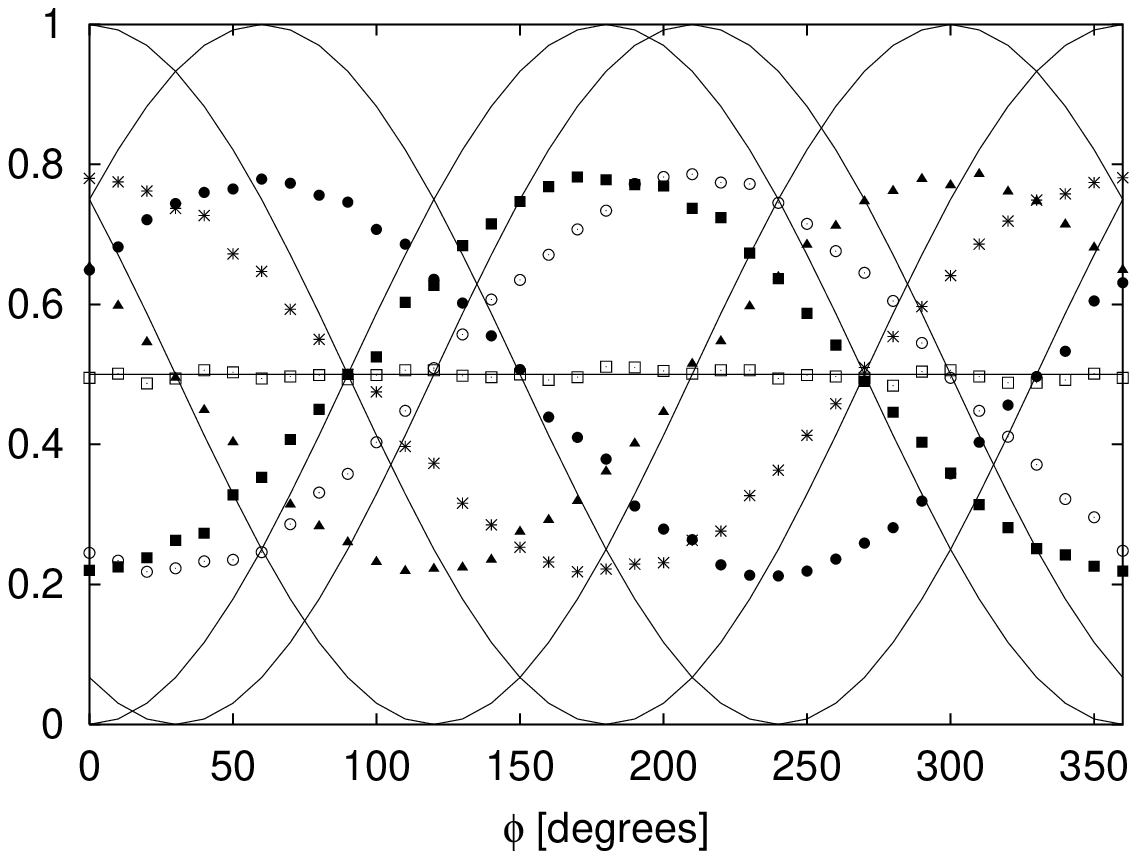}}
\end{picture}
\caption{
Simulation results for the \DLM-network shown in Fig.~\ref{figmz}.
Input channel 0 receives $(\cos\psi_0,\sin\psi_0)$ with probability one.
A uniform random number in the range $[0,360]$ is used to choose the angle $\psi_0$.
Input channel 1 receives no events. %~\cite{GRAN86}.
Each data point represents 10000 events ($N_0+N_1=N_2+N_3=10000$).
Initially the rotation angle $\phi_0=0$ and after each set of 10000 events, $\phi_0$
is increased by $10^\circ$.
Markers give the simulation results for the normalized intensities
as a function of $\phi=\phi_0-\phi_1$.
Open squares: $N_0/(N_0+N_1)$;
Solid squares: $N_2/(N_2+N_3)$ for $\phi_1=0$;
Open circles: $N_2/(N_2+N_3)$ for $\phi_1=30^\circ$;
Bullets: $N_2/(N_2+N_3)$ for $\phi_1=240^\circ$;
Asterisks: $N_3/(N_2+N_3)$ for  $\phi_1=0$;
Solid triangles: $N_3/(N_2+N_3)$ for $\phi_1=300^\circ$.
Lines represent the results of quantum theory (see Eq.~(\ref{MZ3})).
Left: $\alpha=0.98$; Right: $\alpha=0.25$, see the section
on the role of $\alpha$.
}
\label{one-mz}
\end{center}
\end{figure}

In quantum physics~\cite{QuantumTheory}, single-photon experiments with one beam splitter
provide direct evidence for the particle-like behavior of photons~\cite{GRAN86,HOME97}.
The wave mechanical character appears when one performs single-particle
interference experiments.
We now describe a \DLM\ network that displays the same
interference patterns as those observed in single-photon
Mach-Zehnder interferometer experiments~\cite{GRAN86}.
This also proves that the messages generated by the \DLMS\ preserve the phase information
that is essential if the system is to exhibit quantum mechanical behavior.

The schematic layout of the \DLM\ network is shown in Fig.~\ref{figmz}.
Not surprisingly, it is exactly the same as that of a real Mach-Zehnder interferometer.
The \DLM\ network described before is used for the two beam splitters
in Fig.~\ref{figmz}.
The phase shift is taken care of by a passive device that performs a plane rotation.
Clearly there is a one-to-one mapping
from each physical component in the interferometer
to a processing unit in the \DLM\ network.
The processing units in the \DLM\ network
only communicate with each other through the messages (photons)
that propagate through the network.

According to quantum theory~\cite{QuantumTheory}, the amplitudes ($b_0,b_1)$
of the photons in the output modes 0 ($N_2$) and 1 ($N_3$) of the
Mach-Zehnder interferometer are given by~\cite{BAYM74,GRAN86,RARI97,NIEL00}
\begin{eqnarray}
\left(
\begin{array}{c}
b_0\\
b_1
\end{array}
\right)
=
\frac{1}{2}
\left(
\begin{array}{cc}
1&i\\
i&1
\end{array}
\right)
\left(
\begin{array}{cc}
e^{i\phi_0}&0\\
0&e^{i\phi_1}
\end{array}
\right)
\left(
\begin{array}{cc}
1&i\\
i&1
\end{array}
\right)
\left(
\begin{array}{c}
a_0\\
a_1
\end{array}
\right)
.
\label{MZ1}
\end{eqnarray}
Note that in a quantum mechanical setting it is impossible to
simultaneously measure ($N_0/(N_0+N_1)$, $N_1/(N_0+N_1)$)
and ($N_2/(N_0+N_1)$, $N_3/(N_0+N_1)$):
Photon detectors operate by absorbing photons.
In our event-based simulation there is no such problem.

In Fig.~\ref{one-mz} we present a few typical simulation results
for the Mach-Zehnder interferometer built from \DLMS.
We assume that input channel 0 receives
$(\cos\psi_0,\sin\psi_0)$ with probability
one and that input channel 1 receives no events.
This corresponds to $(a_0,a_1)=(\cos\psi_0+i\sin\psi_0,0)$.
We use a uniform random number to determine $\psi_0$.
The data points are the simulation results for the
normalized intensity $N_i/(N_0+N_1)$ for i=0,2,3
as a function of $\phi=\phi_0-\phi_1$.
Lines represent the corresponding results of quantum theory~\cite{QuantumTheory}.
From Fig.~\ref{one-mz} (left) it is clear
that the event-based processing by the \DLM\ network reproduces
the probability distribution
\begin{eqnarray}
|b_0|^2=\sin^2(\frac{\phi_0-\phi_1}{2}),%\nonumber \\
\quad
|b_1|^2=\cos^2(\frac{\phi_0-\phi_1}{2}),
\label{MZ3}
\end{eqnarray}
as obtained from Eq.\Eq{MZ1}.
The results of Fig.~\ref{one-mz} (right) for $\alpha=1/4$
corroborate the analysis of the role of $\alpha$, presented earlier.

\section{Discussion}\label{SUMM}

We have proposed a new procedure to construct algorithms
that can be used to simulate quantum processes
without solving the Schr{\"o}dinger equation.
We have shown
that single-particle quantum interference can be simulated on an event-by-event basis
using local and causal processes, without using concepts such as wave fields
or particle-wave duality.
Our results suggest that we may have discovered
a procedure to simulate quantum phenomena
using causal, local, and event-based processes.
There is a one-to-one correspondence between
the parts of the processing units and network and
the physical parts of the experimental setup.
Only simple geometry is used to construct the simulation algorithm.
In this sense, the simulation approach we propose
satisfies Einstein's criteria of realism and causality~\cite{HOME97}.
Our approach is not an extension of quantum theory in any sense
and is not a proposal for another interpretation of quantum mechanics.
The probability distributions of quantum theory are
generated by a particle-like, causal learning process,
and not vice versa~\cite{PENR90}.

%\section*{Acknowledgement}
%We thank S. Miyashita for extensive discussions.

%Support from the ``Nederlandse Stichting Nationale Computer Faciliteiten (NCF)'' is gratefully acknowledged.

%\ifDONOTSHOW

\raggedright


\begin{thebibliography}{999}
\bibitem{GRAN86}
P. Grangier, R. Roger, and A. Aspect,
Europhys. Lett. {\bf 1}, 173 (1986)
%
\bibitem{TONO98}
A. Tonomura,
{\it The Quantum World Unveiled by Electron Waves},
World Scientific, Singapore (1998)

\bibitem{PerfectExperiments}
In this paper we disregard
limitations of real experiments such as detector efficiency,
imperfection of the source, and the like.

\bibitem{BALL03}
L.E. Ballentine,
{\it Quantum Mechanics: A Modern Development},
World Scientific, Singapore (2003)

\bibitem{HOME97}
D. Home, {\it Conceptual Foundations of Quantum Physics},
Plenum Press, New York (1997)

\bibitem{FEYN65}
R.P. Feynman, R.B. Leighton, M. Sands,
{\it The Feynman lectures on Physics}, Vol. 3,
Addison-Wesley, Reading MA (1996)

\bibitem{PENR90}
R. Penrose, {\it The Emperor's New Mind},
Oxford University Press, Oxford (1990)

\bibitem{KAMP88}
N.G. Van Kampen,
Physica A {\bf 153}, 97 (1988)

\bibitem{QuantumTheory}
We make a distinction between quantum theory and quantum physics.
We use the term {\sl quantum theory} when we refer to
the mathematical formalism, i.e., the postulates
of quantum mechanics (with or without the wave function
collapse postulate)~\cite{BALL03} and the rules (algorithms) to compute the
wave function.
The term {\sl quantum physics} is used for microscopic, experimentally observable phenomena
that do not find an explanation within the mathematical
framework of classical mechanics.

\bibitem{HAYK86}
S. Haykin,
{\it Adaptive Filter Theory},
Prentice Hall, New Jersey (1986)

\bibitem{KOEN04}
K. De Raedt, H. De Raedt, and K. Michielsen,
{\sl Deterministic event-based simulation of quantum interference},
arXiv: quant-ph/0409213

\bibitem{HANS05}
H. De Raedt, K. De Raedt, and K. Michielsen,
{\sl New method to simulate quantum interference using deterministic processes
and application to event-based simulation of quantum computation},
to appear in J. Phys. Soc. Jpn.

\bibitem{KRIS04}
K. Michielsen, K. De Raedt, and H. De Raedt,
{\sl Simulation of Quantum Computation: A deterministic event-based approach},
to appear in J. Comp. Theor. Nanoscience

\bibitem{MZIdemo}
An interactive program that performs the event-based simulations
of a beam splitter, one Mach-Zehnder interferometer, and
two chained Mach-Zehnder interferometers can be found at
http://www.compphys.net/dlm

\bibitem{HAYK99}
S. Haykin,
{\it Neural Networks},
Prentice Hall, New Jersey (1999)

\bibitem{BAYM74}
G. Baym,
{\it Lectures on Quantum Mechanics},
W.A. Benjamin, Reading MA (1974)

\bibitem{RARI97}
J.G. Rarity and P.R. Tapster,
Phil. Trans. R. Soc. Lond. A {\bf 355}, 2267 (1997)

\bibitem{NIEL00}
M. Nielsen and I. Chuang,
{\it Quantum Computation and Quantum Information},
Cambridge University Press, Cambridge (2000)
%
\end{thebibliography}
\end{document}